\documentclass{aa}
\usepackage{epsf}
\usepackage{graphicx}

\newcommand{\magpt}[2]{\mbox{$\rm #1\hspace{-0.25em}\stackrel{m}{.}
      \hspace{-1.0mm}#2$}}                             

\newcommand{\halpha}{H$_\alpha$}

\newcommand\teff{T_{\mathrm{eff}}}

\newcommand{\rsun}{R_\odot}
\newcommand{\msun}{M_\odot}
\newcommand{\kms}{km~s$^{-1}$}
\newcommand{\hochth}{$^{\mathrm{th}}$}

\begin{document}
\title{Binaries discovered by the SPY project. II. HE\,1414--0848: a 
double degenerate with a mass close to the
Chandrasekhar limit\thanks
{Based on data obtained at the Paranal Observatory of the European 
Southern Observatory for programs No. 165.H-0588, 266.D-5658, and 167.D-0407}
}
\author{R.~Napiwotzki\inst{1}, 
D.~Koester\inst{2}, 
G.~Nelemans\inst{3}, 
L.~Yungelson\inst{4},
N.~Christlieb\inst{5},
A.~Renzini\inst{6},
D.~Reimers\inst{5}, 
H.~Drechsel\inst{1}, 
\and B.~Leibundgut\inst{6}
}
\offprints{R. Napiwotzki}
\institute {Dr. Remeis-Sternwarte, Astronomisches Institut der Universit\"at
Erlangen-N\"urnberg, Sternwartstr. 7, D-96049 Bamberg, Germany; e-mail: napiwotzki@sternwarte.uni-erlangen.de
\and
Institut f\"ur Theoretische Physik und Astrophysik, 
Universit\"at Kiel, D-24098 Kiel, Germany
\and
Institute of Astronomy, Madingley Road, CB3~0HA, Cambridge, UK
\and
Institute of Astronomy of the Russian Academy of Sciences, 48 
Pyatnitskaya Str., 109017 Moscow, Russia
\and
Hamburger Sternwarte, Universit\"at Hamburg, Gojenbergsweg 112, 
D-21029 Hamburg, Germany
\and
European Southern Observatory, Karl-Schwarzschild-Str. 2,
D-85748 Garching, Germany
}

\date{}

\abstract{In the course of our search for double degenerate (DD) binaries
as potential progenitors of type Ia supernovae with the UVES
spectrograph at the ESO VLT (ESO {\bf S}N\,Ia {\bf P}rogenitor
surve{\bf Y} -- SPY) we discovered that the white dwarf
HE\,1414$-$0848 is a double-lined DA+DA binary with an orbital period of
$P = 12^{\mathrm{h}} 25^{\mathrm{m}} 39^{\mathrm{s}}$. 
Semi-amplitudes of 128\,km\,s$^{-1}$ and 100\,km\,s$^{-1}$ are derived
for the individual components.  The amplitude ratio and the measured
difference in gravitational redshift is used to estimate the masses of
the individual components: $0.55M_\odot$ and $0.71M_\odot$. Hence the
total mass of the HE\,1414$-$0848 system is $1.26M_\odot$, only 10\%
below the Chandrasekhar limit.  The results of a model atmosphere
analysis are consistent with our mass estimated from the
orbit. Temperatures of the individual components are also determined. 
Possible scenarios for the formation of this system are discussed.
The system will merge due to loss of
angular momentum via gravitational wave radiation after two Hubble
times.  HE\,1414$-$0848 does not qualify as a SN\,Ia
progenitor, but it is the most massive close DD known
today.
}

\authorrunning{Napiwotzki et al.}
\titlerunning{Binaries discovered by the SPY project. II. HE\,1414-0848}
\maketitle

\section{Introduction\label{intro}} 
 
Supernovae of type Ia (SN\,Ia) play an outstanding role in the study of
cosmic evolution.
In particular they are regarded as one of the best standard candles to
determine the cosmological parameters $H_\circ$, $\Omega$ and
$\Lambda$ (e.g.\ Riess et al.\ \cite{ries98}, Leibundgut \cite{L01}).  
However, the nature of
their progenitors remains a mystery (e.g. Livio \cite{liv00}).

There is general consensus that the SN event is due to the
thermonuclear explosion of a white dwarf (WD) when a critical mass
(likely the Chandrasekhar limit) is reached, but the nature of the
progenitor system remains unclear.  While it must be a binary, with
matter being transferred to the WD from a companion until the critical
mass is reached, two main options exist for the nature of the mass
donator: either another WD in the so-called double degenerate (DD)
scenario (Iben \& Tutukov \cite{ibtu84}), or a red giant/subgiant in the
so-called single degenerate (SD) scenario (Whelan \& Iben \cite{whel73}).

As progenitor candidate, the 
DD model considers a binary with the total mass of the white dwarf
components larger than the Chandrasekhar mass, which merges in less
than a Hubble time due to the loss of angular momentum via gravitational
wave radiation.  Several systematic radial velocity (RV) searches for
DDs have already been undertaken starting in the mid 1980's (see Marsh
\cite{mar00} for a review).  By now, combining all the surveys $\approx
180$ WDs have been checked for RV variations yielding a sample of 18 DDs with
$P<6\fd 3$ (Marsh \cite{mar00}; Maxted et al.\ \cite{MMM00}).
Six of the 18 known DDs are double-lined systems (in three of these 
systems the 
companion is barely detectable). 
None of the 18 known DD systems seems massive enough to qualify as a SN~Ia
precursor. This is not surprising, as theoretical simulations suggest
that only a few percent of all close DDs are potential SN\,Ia progenitors
(Iben et al.\ \cite{ITY97}; Nelemans et al.\ \cite{NYP01}).
Recently, the binary KPD\,1930+2752, consisting of a subluminous B (sdB) star
and an invisible white dwarf component, was proposed as potential SN\,Ia
progenitor by Maxted et al.\ (\cite{MMN00}). The system mass
exceeds the Chandrasekhar limit, and the system will merge within a
Hubble time. However, it is not clear, if the merger event will produce a
SN\,Ia explosion (Ergma et al.\ \cite{EFY01}).

In order to perform a definite test of the DD scenario we have embarked on a 
large spectroscopic survey of 1500 white dwarfs using the UVES spectrograph at 
the ESO VLT UT2 (Kueyen) to search for white 
dwarfs and pre-white dwarfs with variable RVs 
(ESO {\bf S}N\,Ia {\bf P}rogenitor surve{\bf Y}
-- SPY). 
The SPY project yields a wealth of new RV variable DDs (Napiwotzki et al.\
\cite{NCD02}). An analysis of
the subdwarf B + white dwarf system HE\,1047$-$0436 was presented in the 
first paper of this series (Napiwotzki et al.\ \cite{NEH01}).

Among the newly detected DDs are six new double-lined binaries.
In all systems both WD components can be easily
recognised.  Here we report on follow-up spectroscopy of
HE\,1414$-$0848 (cf.\ Sect.~2). The determination of the radial velocity
curves, orbital parameters, and masses for both
components is described in Sect.~3.  A model atmosphere analysis is
carried out and the results are compared with the analysis of the
RV curves in Sect.~4.  We finish with a discussion
of the HE\,1414$-$0848 system and possible formation scenarios.

\section{Observations and Data Analysis\label{obs}}

HE\,1414$-$0848 ($\alpha_{2000}=14^{\mathrm{h}} 16^{\mathrm{m}}
52.1^{\mathrm{s}}$, $\delta_{2000}=-9^{\circ} 02'04''$,
$B_{\mathrm{pg}}=$\magpt{16}{2}) was discovered by the Hamburg ESO
survey (HES; Wisotzki et al.\ \cite{wis00}, Christlieb et al.\
\cite{chri01}) as a potential cool white dwarf and, therefore, was
included in our survey.  The survey spectra showed that it is a
double-lined system consisting of two DA white dwarfs. Both components
were visible in the \halpha\ core separated by 4.6\,\AA, corresponding
to 210\,km/s, in both discovery spectra. This indicated high orbital
velocity and a short period, making HE\,1414$-$0848 a high-priority
target for follow-up studies.

Fifteen high resolution (0.3\AA) echelle spectra of
HE\,1414$-$0848 have been secured with VLT-UVES between March 9 and
April 5, 2001 in service mode.  Details on the observational set up of the
UVES instrument and the data reduction can be found in Koester et
al. (\cite{KNC01}) and Napiwotzki et al.\ (\cite{NCD02}). 
The chosen observing strategy was to start with
several closely spaced exposures during the first night. This should
provide us with a rough first estimate of the period and radial
velocity curve, which could be used to extrapolate the orbit to the
next nights. Consequently, a decreasing number of spectra was taken
during the following nights, finishing with single exposures separated
by more than one day.  This strategy in conjunction with the
high quality of the UVES data allowed the determination of accurate orbital
parameters with a relatively small number of spectra.

\begin{figure}
\epsfxsize8.5cm
\epsffile[50  40  590 745]{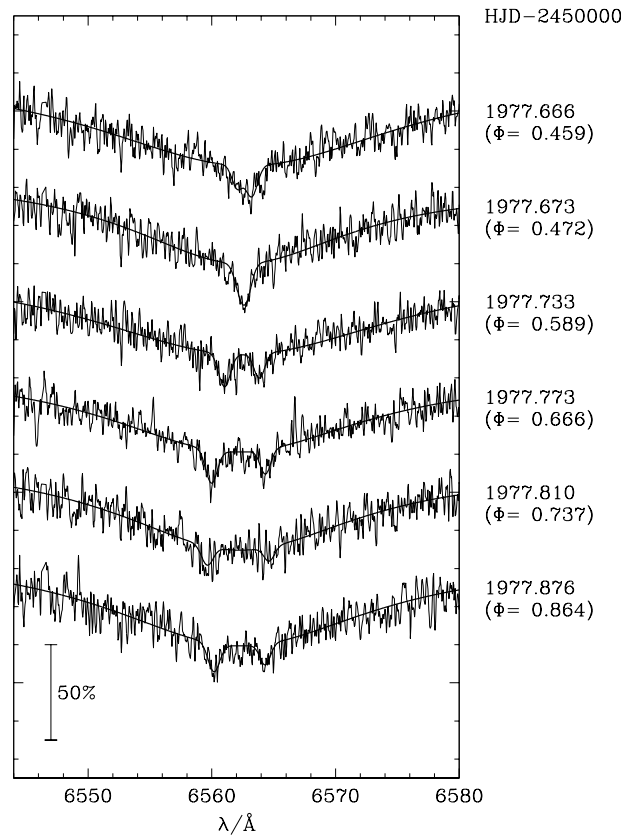}
\caption{\halpha\ region of the spectra covering a timespan of 
5$^{\mathrm{h}}$ in the night of
March 8/9, 2001 together with the fits described in Sect.~\ref{s:RV}. 
The numbers indicate the Julian date of the exposures and the orbital phase
$\phi$ (cf.\ Fig.~\ref{f:RV}). Star ``A''
in the text corresponds to the blueward shifted line core in, e.g.,
the JD 2451977.773 exposure, ``B'' to the red component. Observed spectra 
were smoothed with a three pixel boxcar (0.09\,\AA).}
\label{f:halpha}
\end{figure}
 
Six \halpha\ spectra taken during the first night of the 
follow-up observations are displayed in Fig.~\ref{f:halpha} together
with a fit of both components as described in Sect.~\ref{s:RV}. The rapid
change of the spectral appearance due to the orbital motion is obvious.   

\section{Radial velocity curve and orbital parameters\label{s:RV}}

Although the line cores of both components are similar, the left one in
Fig.~\ref{f:halpha} is slightly deeper and broader (called component A
further on). We used this in a
first step to identify both stars in each spectrum. In a second step
we fitted the central region ($\pm 20$\,\AA)
by using two Gaussians (one for every
component), a Lorentzian to model the line wings (parameters of the Lorentzian
were determined from a fit covering $\pm 50$\,\AA\ and then held fixed), 
and a linear polynomial
to reproduce the overall spectral trend. 
Fitting of the H$_\alpha$ profiles was performed with a downhill simplex
algorithm, which will be described in a forthcoming paper.

\begin{table}
\caption{Fitted heliocentric radial velocities. Hel.JD was calculated for the
middle of the exposures.\label{t:RVs}}
\begin{tabular}{r|r@{$\pm$}lr@{$\pm$}l}
\multicolumn{1}{c|}{Hel.JD}		
	&\multicolumn{4}{c}{Heliocentric RV} 
	\\ 
$-$2,400,000		&\multicolumn{2}{c}{ A}
	&\multicolumn{2}{c}{ B}	
	\\ \hline
51684.68664  &$-$110&4	&98&4		\\
51687.61061  &129&4	&$-$80&4	\\
51977.66606  &47&9 	&$-$3&11 	\\
51977.67313  &24&4  	&$-$4&6		\\
51977.73344  &$-$59&5 	&70&5		\\
51977.77310  &$-$106&4	&92&5		\\
51977.81012  &$-$119&9	&104&11		\\
51977.87602  &$-$98&5	&91&6		\\
51978.67757  &67&4	&$-$34&6	\\
51978.75510  &$-$50&4	&56&4		\\
51978.85895  &$-$124&5	&112&5		\\
51979.68326  &92&8	&$-$64&9	\\
51979.88119  &$-$123&7	&113&8		\\
51987.70636  &$-$95&3	&92&5		\\
52003.59657  &$-$29&3	&39&4		\\
52003.82394  &$-$25&3	&44&3		\\
52004.58719  &41&4	&$-$20&5	\\
52137.53531  &127&3	&$-$87&4	\\
\end{tabular}
\end{table}

\begin{figure}
\epsfxsize8.6cm
\epsffile[5 15 740 540]{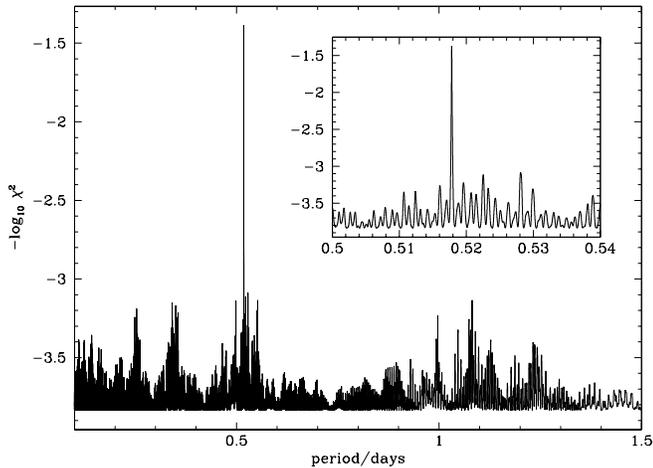}
\caption{Power spectrum of the RV measurements of component A. The inset shows
details of the region around the main peak.
\label{f:power}}
\end{figure}

In a third
step we fitted RV curves for a range of periods and produced a ``power
spectrum'' indicating the fit quality as a function of period 
(like the one presented in
Fig.~\ref{f:power}). In a fourth step we inspected the phased RV curves
using periods corresponding to peaks in the power spectrum and 
compared them with the measured
values. Most could be ruled out, because of discrepant RV measurements.
Finally, one unambiguous solution remained. 

\begin{table}
\caption{Orbital parameters of the HE\,1414$-$0848 system. Given are the
``system'' velocities $\gamma_0$, RV amplitudes $K_{\mathrm{A/B}}$, the
$\chi^2$ values of the fits and the number of data points $N$. 
\label{t:orbit}}
\begin{tabular}{l|rr|cc}
Comp.\	&$\gamma_0$ [km/s]	&$K_{\mathrm{A/B}}$ [km/s] &N &$\chi^2$	
\\ \hline
A	&$2.1\pm 1.1$	&$128.2\pm 1.6$	 &18 &23.7\\ 
B	&$16.4\pm 1.3$	&$99.9\pm 1.7$ 	 &18 &25.2\\
\end{tabular}
\end{table}

\begin{figure}
\epsfxsize8.6cm
\epsffile[5 15 740 540]{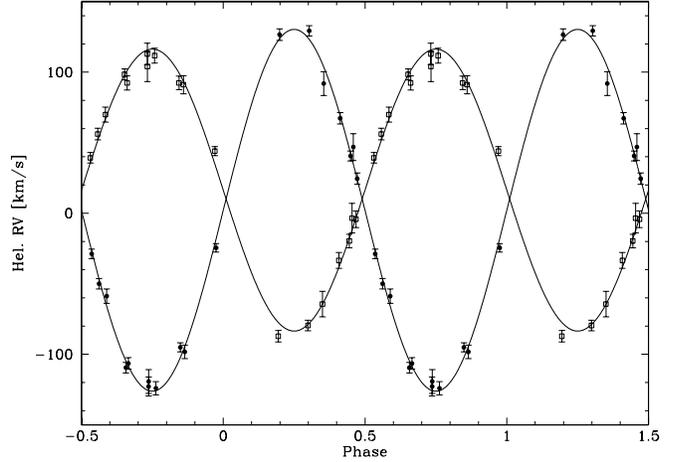}
\caption{Measured radial velocities as a function of orbital phase and 
fitted sine curves for HE\,1414-0848. Filled circles correspond to 
component ``A'' and open rectangles to ``B''. Note the difference of the 
``systemic velocities'' $\gamma_0$ between both components caused by the
gravitational redshift.
\label{f:RV}}
\end{figure}

In the fifth and last step we extrapolated the derived RV curves back
to the time of the discovery observations and to an additional spectrum taken
in August 2001. We used the phase information
to identify the
components in these exposures, and added the RVs to our analysis.
Individual measurements are listed in Table~\ref{t:RVs}.
Since the spectra span more than one year, 
a very accurate period could be determined:
$P = 12^{\mathrm{h}} 25^{\mathrm{m}} 39^{\mathrm{s}}$.
Results for both components agree within 1\,s.
Semi-amplitudes $K_{\mathrm{A/B}}$ and the ``system velocities'' 
$\gamma_0$ 
are given in Table~\ref{t:orbit}.
Accordingly the ephemeris for the time $T_0$ defined as the conjunction time at
which star A moves from the blue side to the red side of the RV curve 
(i.e.\ star A is closest to the observer)  is
\begin{equation}
\begin{array}{lrrl}
\mathrm{Hel.JD}(T_0) = &2451976.911& +\: 0.51781 &\times E\\
			& \pm 0.001& \pm 0.00001&  \\
\end{array}\,.
\end{equation}
Error limits are given in the second line.
The mass ratio of both white dwarfs can be computed from the ratio of
RV amplitude:
\begin{equation}
\frac{M_{\mathrm{B}}}{M_{\mathrm{A}}} = 
\frac{K_{\mathrm{A}}}{K_{\mathrm{B}}} = 1.28 \pm 0.02\,.
\label{eq:Mrel}
\end{equation}
From Fig.~\ref{f:RV} and Table~\ref{t:orbit} it is evident that the
``system velocities'' $\gamma_0$ derived for component A and B differ 
much more than naively expected considering the error bars:
$\gamma_0(\mathrm{B}) -\gamma_0(\mathrm{A}) = 14.3\pm 1.7$\,km/s . However,
this is easily explained by the mass dependent gravitational redshift
of white dwarfs
\begin{equation}
z = \frac{GM}{Rc^2}\,.
\end{equation}
This offers the opportunity to determine masses of the individual
white dwarfs in the HE\,1414$-$0848 binary. For a given mass-radius
relation gravitational redshift can be computed as a function of white
dwarf mass.  Since the mass ratio is given by Eq.~\ref{eq:Mrel}, only
one combination of masses can fulfil both constraints.
Since HE\,1414$-$0848 is the outcome of common envelope evolution, it is not
clear which mass-radius relation to use. We performed the analysis
with mass-radius relations constructed for $\teff = 9800$\,K (cf.\
Sect.~\ref{s:analysis}) from three different white dwarf cooling
tracks, which should give us an estimate of systematic errors
introduced by a particular choice.  We used the tracks computed by
Bl\"ocker et al.\ (\cite{BHD97}), which result from the self-consistent
calculation of single star evolution and cooling curves computed by
Wood (\cite{Woo95}) for white dwarfs with a ``thick'' hydrogen layer
($M_{\mathrm{H}} = 10^{-4}M_{\mathrm{WD}}$) and a ``thin'' layer
($M_{\mathrm{H}} = 0$).  The resulting white dwarf masses are
$M_{\mathrm{A}} = 0.55\pm 0.03 M_\odot$ and 
$M_{\mathrm{B}} = 0.71\pm 0.03M_\odot$
for Wood's models with ($M_{\mathrm{H}} = 0$). The resulting masses
are $0.01M_\odot$ larger/lower if we adopt the ($M_{\mathrm{H}} =
10^{-4}M_{\mathrm{WD}}$) models or the tracks of
Bl\"ocker et al.\ (\cite{BHD97}), respectively.  Thus the choice of a
particular mass-radius relation is of little importance.
The mass sum of both white dwarfs is $M = 1.26\pm 0.06M_\odot$ (errors
propagated from the orbital parameters + possible systematic errors).
Thus HE\,1414$-$0848 is the most massive DD ever found with a total
mass only 10\% below the Chandrasekhar limit!

The separation between both white dwarfs is quite small, only $2.9R_\odot$.
Thus HE\,1414$-$0848 obviously underwent phases of strong binary 
interaction in its history. From the size of the orbits and the period orbital
velocities can be computed: 164\,\kms\ for component A and 123\,\kms\ for B. 
The comparison with the observed RV amplitudes allows us to determine the
inclination of this system as $i = 52^{\circ}$.

\section{Spectroscopic analysis\label{s:analysis}}

\subsection{Spectral fitting with one model}

Further insight into the HE\,1414$-$0848 system can be gained from a
model atmosphere analysis. Since this system is double-lined the
spectra are a superposition of both individual white dwarf spectra. A
deconvolution is beyond the scope of this paper, but an analysis of
the combined spectrum will already allow us to derive constraints on
this system. We used four spectra obtained close to conjunction for
this purpose (Table~\ref{t:fit}). Radial velocity differences are
small and do not cause significant artificial broadening of the line
profiles.

\begin{figure}
\epsfxsize8cm
\epsffile{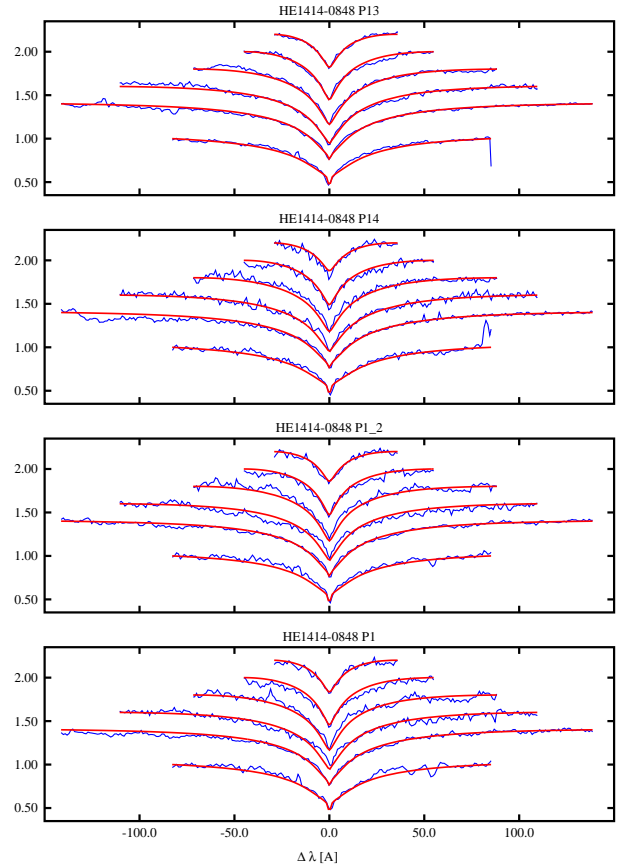}
\caption{Fit of four spectra of HE\,1414$-$0848 taken close to
conjunction (cf.\ Table~\ref{t:fit}).
\label{f:fit}}
\end{figure}

The spectra were rebinned to a resolution of approximately 1\,\AA.
A large grid of LTE DA models computed with a model atmosphere code
described in Finley et al.\ (\cite{FKB97}) was used for the analysis. 
A simultaneous fit of the Balmer lines H$\alpha$ to H$_8$ using a
$\chi^2$ minimisation technique was performed. For details refer to
Koester et al.\ (\cite{KNC01} and references therein). The 
fits are shown in Fig.~\ref{f:fit} and individual results are given 
in Table~\ref{t:fit}. The average parameters are $\teff = 9800\pm 90$\,K and 
$\log g = 8.17\pm 0.09$ (errors correspond to the scatter of individual
measurements). 

\begin{table}
\caption{Orbital phases $\phi$ and fit results for the spectra 
displayed in Fig.~\ref{f:fit}. 
\label{t:fit}}
\begin{tabular}{l|r|ll}
spectrum	&\multicolumn{1}{c|}{$\phi$}
	 &\multicolumn{1}{c}{$\teff$ [K]} &$\log g$ [cm/s$^2$] \\ \hline
P1	&0.459 &$9739\pm 29$  &$8.12\pm 0.04$ \\
P1\_2	&0.472 &$9722\pm 27$  &$8.21\pm 0.04$ \\
P13	&0.975 &$9924\pm 26$  &$8.29\pm 0.04$ \\
P14	&0.449 &$9822\pm 19$  &$8.08\pm 0.02$ 
\end{tabular}
\end{table}

Since the observed spectra are a combination of spectra of both components,
the interpretation of this result is not straightforward. However, we expect
that the analysis results represent some sort of average of the 
individual values of both components (especially, if both components are 
similar as in the case of HE\,1414$-$0848; cf.\ next Section). 
We used Wood's (\cite{Woo95}) mass radius-relation for ($M_{\mathrm{H}}=0$)
to estimate a mass of $0.68\pm 0.06 M_\odot$ for these parameters.
This is consistent with the results of our analysis of the RV curve.

\subsection{Spectral fitting with two models}

\begin{figure}
\epsfxsize8cm
\epsffile{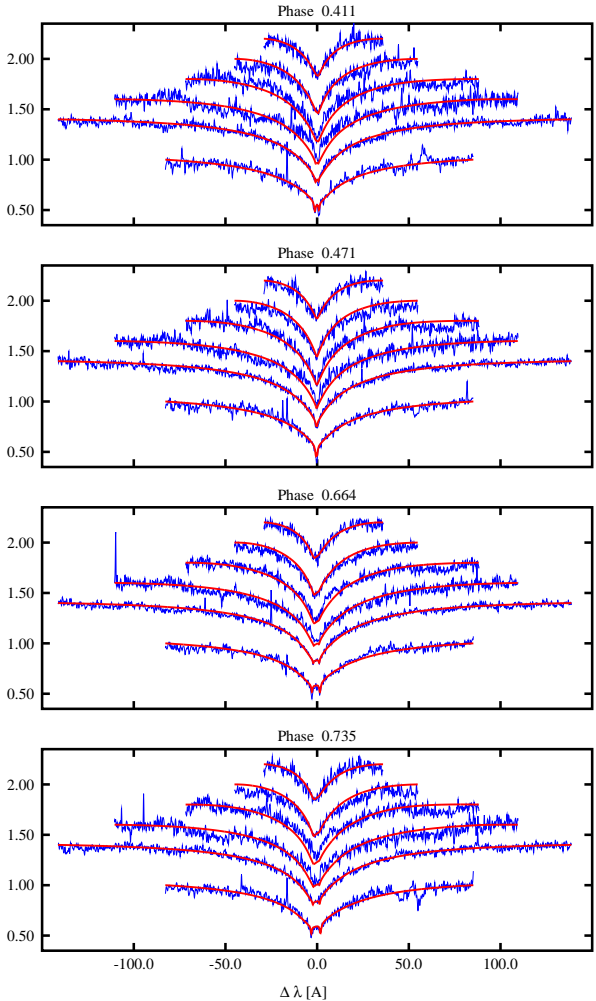}
\caption{Combination of the best fitting model spectra for components A and B
($\teff(\mathrm{A})=8900$\,K and $\teff(\mathrm{B})=10800$\,K
and four different phases of the binary orbit.
\label{f:doublefit}}
\end{figure}

Fitting the combined spectra of the two white dwarfs can only give an
approximate average of the atmospheric parameters. We have therefore
tried to use 14 individual spectra covering different phases of the
orbit and fit them with a combination of two DA spectra (another
spectrum -- 12 -- has a very perturbed H$\alpha$ profile and could not
be used). The quality of these spectra varies and the total number of
fit parameters is high -- it is not possible to determine them
unambiguously. We have therefore used the radial velocities of the
individual components and their masses from the radial velocity curve
and held them fixed during the fitting. Surface gravities are 8.18 and
7.87; this also determines the relative weight of the two model
spectra from the radius, obtained from the mass-radius relation of
Wood (1995); the weight ratio is 40:60. The only remaining parameters
are then the two effective temperatures of the models, which are
obtained through a $\chi^2$ fitting routine similar to the methods
described in Koester et al. (2001).

Table~\ref{t:doublefit} gives the results for two different attempts. 
With 
Method~1 we have used only the inner parts of H$\alpha$ and H$\beta$ (80 -
100 \AA) to determine the atmospheric parameters. In Method~2 we performed a
simultaneous fit of all Balmer lines from H$\alpha$ to H$_8$.

This exercise demonstrates that the individual temperatures are
similar for both components and are in fact fairly close to the
values obtained from fitting the combined spectra with one average
model. However, the scatter from the different spectra is quite
large -- even the assignment, which of the two stars is the hotter one
is not the same in all phases. Obviously the demand on the
signal-to-noise ratio and on the calibration is higher for this method
than for fitting single stars, especially if the two spectra are very
similar as in this case. 

The most meaningful results can be expected
during the quadrature phases ($\phi = 0.25$ and 0.75), when the RV
separation between both components is largest. Thus not unexpectedly
the scatter of the temperature determinations for the corresponding
spectra P3, P4, P5, P6, and P10 (marked boldface in
Table~\ref{t:doublefit}) is relatively small. Our best temperature
estimates are $8900\pm 550$\,K for component A and $10790\pm 550$\,K
for B from Method~2 ($9330\pm 640$\,K and $10310\pm 480$\,K if
Method~1 is adopted). The error margins correspond to the scatter of
individual measurements. Sample spectra 
fitted with these parameters for several orbital phases are displayed 
in Fig.~\ref{f:doublefit}.

\begin{table*}
\caption{Results for the effective temperatures and their 1$\sigma$
formal errors of the two DA components from spectra of different
phases. Index (B) indicates the more massive star. Columns 3-6 correspond
to 
Method~1, where only the inner parts of H$\alpha$ and H$\beta$ were used in
the fit. Columns 7-10 correspond to Method~2, which used H$\alpha$ to H$_8$.}
\label{t:doublefit}
\begin{center}
\begin{tabular}{|l|r||r|r|r|r||r|r|r|r|}
\hline
spectrum &$\phi$ &$\teff(\mathrm{B})$& $\sigma(\mathrm{B})$& 
$\teff(\mathrm{A})$& $\sigma(\mathrm{A})$& 
$\teff(\mathrm{B})$& $\sigma(\mathrm{B})$& 
$\teff(\mathrm{A})$& $\sigma(\mathrm{A})$\\
\hline
P1   &0.458&  8614 &  76 & 10143 &  20 &  10065 &  78 &  9427
&  60\\
P1\_2  &0.471&  9668 & 210 &  9812 & 133 &   9819 & 204 &  9735 & 131\\
P2    &0.581&  9950 & 102 &  9345 &  88 &   7837 & 105 &  9974 &  12\\
{\bf P3}    &0.664&  9811 & 148 &  9756 &  99 &  11076 &  18 &  8612 &  28\\
{\bf P4}    &0.736& 10867 &  22 &  8482 &  41 &  10916 &  22 &  8329 &  38\\
{\bf P5}    &0.863& 10769 &  32 &  8793 &  38 &   9814 & 113 &  9798 &  72\\
{\bf P6}    &0.847& 10006 & 141 &  9900 &  87 &  11056 &  31 &  9154 &  24\\
P7    &0.534&  9748 & 171 &  9858 & 109 &   9807 & 171 &  9803 & 110\\
P8    &0.411&  9303 & 111 & 10161 &  48 &  10198 &  69 &  9411 &  59\\
P9    &0.561&  8087 &  72 & 10219 &   9 &  10862 &  20 &  8574 &  32\\
{\bf P10}   &0.761& 10100 & 139 &  9696 &  99 &  11075 &  20 &  8611 &  31\\
P11   &0.353&  9765 & 267 &  9728 & 172 &  11087 &  37 &  7648 &  81\\
P13   &0.973&  9722 & 131 &  9823 &  83 &  10151 &  62 &  9683 &  45\\
P14   &0.447& 10005 & 219 &  9900 & 137 &   9441 & 111 & 10210 &  48\\
\hline
\end{tabular}
\end{center}
\end{table*}

\section{Discussion
}

From the above derived temperatures we can estimate the ages of both 
components using cooling tracks for white dwarfs. 
Ages were interpolated  from the tracks of Wood ({\cite{Woo95}) for
the masses of the white dwarfs (temperatures from Method~2 adopted). 
Resulting ages are $(1.2\pm 0.2)\cdot 10^9$\,yrs for A and
$(1.0\pm 0.3)\cdot 10^9$\,yrs for B.
Thus the cooling ages of both white dwarfs are very similar and although B is
$\approx$2000\,K hotter than A it could, within our error margins, 
have been formed first.

\begin{figure} 
\epsfxsize8cm
\epsffile[40 32 701 480]{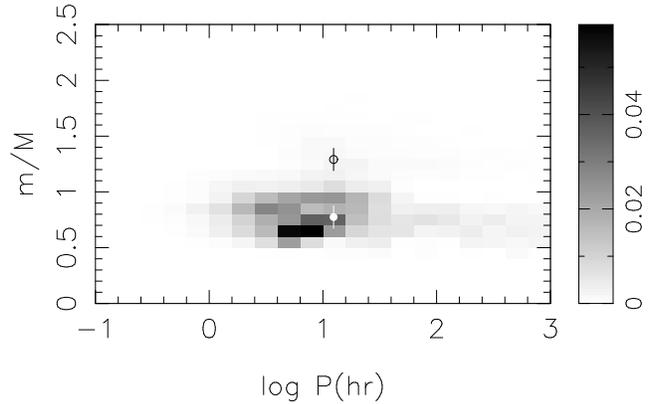}
\caption[]{Period-mass distribution for the current population of DDs in
	the Galaxy for systems in which at least one white dwarf has a
	mass above $0.7M_\odot$ from the best model (A3) of Nelemans et al.\ 
	(\cite{NYP01}). Grey shades indicate the expected number of systems 
	brighter than $V=15^{\mathrm{m}}$. 
	The symbols show the possible mass ratios'
	for HE\,1414$-$0848 depending on which of the two components
	was formed first.}
\label{f:pq}
\end{figure}

We can compare the properties of HE\,1414$-$0848 with the outcome of 
computations for the synthesis of the population of
double white dwarfs of (Nelemans et al.\ \cite{NYP01}). 
In Fig.~\ref{f:pq} we compare the predicted distribution over period and mass
ratios' for systems with masses similar to the masses of HE\,1414$-$0848.
The mass ratio in this plot $m/M$ is defined as the mass of the last
formed white dwarf over the mass of the first formed white dwarf. Since in
the case of HE\,1414$-$0848 it is not reliable known which white dwarf
formed first, we plotted both possible mass ratios, 1.28 and 0.78, in
Fig.~\ref{f:pq}. The population synthesis indicates a larger probability
for the systems with $m/M=0.78$. In Fig.~\ref{f:scenario} we give an
example of a scenario leading to such a DD system.

A binary with an initial semi-major
axis of $\approx$$150 \rsun$ has components with initial
masses of $\approx$4 and $3 \msun$. The primary fills its Roche
lobe in the AGB stage and unstable mass transfer ensues. We estimate
the outcome of the mass transfer using an angular momentum balance
(Nelemans et al.\ \cite{NVY}). The orbit of the system shrinks slightly and
the primary becomes a hot carbon-oxygen (CO) white dwarf. After 
$\approx$170 Myr the secondary fills its Roche lobe as an AGB star.  The mass
transfer is again unstable, the common envelope is lost, and a close
binary consisting of two CO white dwarfs, just like HE1414-0848, is the
result (see Fig~\ref{f:scenario}). The period of the system decreases
during the subsequent evolution due to angular momentum loss via
radiation of gravitational waves.

\begin{figure} 
\epsfxsize8cm
\epsffile[185 62 701 521]{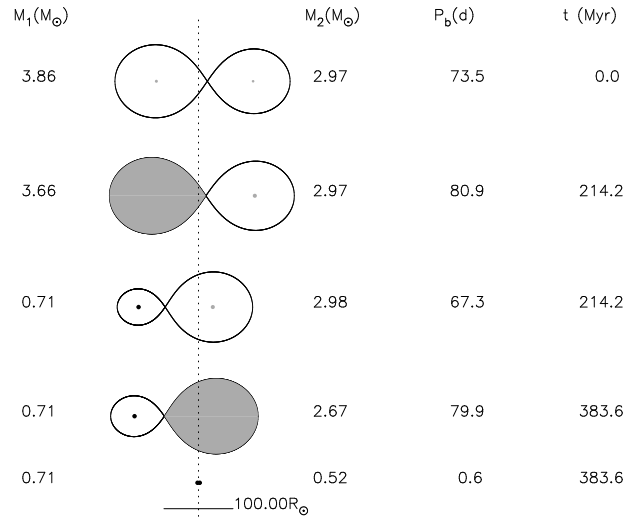}
\caption[]{A possible scenario
  for the formation of a system similar to HE1414-0848, 
  starting with a $3.86 + 2.97 \msun$
  star in a 73.5 day orbit. The black line shows the Roche lobe for
  both stars, filled grey if the star fills its Roche lobe. The ruler
  at the bottom shows the scale.}
\label{f:scenario}
\end{figure}

\section{Conclusions}

We report the discovery of the double-lined radial velocity variable 
binary HE\,1414$-$0848.
We measured accurate radial velocities for both components 
from high resolution spectra and derived an orbital period of 
$12^{\mathrm{h}}25^{\mathrm{m}}39^{\mathrm{s}}$ and semi-amplitudes
of 128\,km/s and 100\,km/s, respectively. We combined the mass ratio 
computed from this amplitudes and the measured difference of gravitational
redshifts and determined masses of $0.55\pm 0.03M_\odot$ and 
$0.71\pm 0.03M_\odot$ for
the components using a mass-radius relation.  
This is the first application of this method to a DD system.
The total mass of the HE\,1414$-$0848 system is $1.26\pm 0.06 M_\odot$, only
10\% below the Chandrasekhar limit,  making it the most massive DD ever 
found in a RV survey. It will merge within two Hubble times (24\,Gyrs).

We performed spectral fitting of both individual components and derived 
$\teff = 10800\pm 550$\,K for the more massive component B and $\teff = 
8900\pm 550$\,K for component A. Ages can be estimated from the WD cooling 
tracks of Wood ({\cite{Woo95}). The resulting cooling ages for both 
white dwarfs are close to 1\,Gyr.  

We compared the resulting orbital parameters of HE\,1414$-$0848 with the
outcome of theoretical calculations for the formation of DDs and find 
that HE\,1414$-$0848 seems to be a fairly typical double white dwarf in terms
the masses of the two white dwarfs and its orbital period.
 
\acknowledgement 
{We express our gratitude to the ESO staff, for providing invaluable help 
and conducting the service observations and pipeline reductions, which have 
made this work possible. 
This project is supported by the DFG (grant Na365/2-1). L.Y.\ acknowledges 
support by RFBR and ``Program Astronomy'' grants
and warm hospitality and support of the Astronomical Institute
``Anton Pannekoek''. 
}

\end{document}